\documentstyle[twocolumn,prl,aps]{revtex}

\begin{document}
\draft

\title{Low-lying Quasiparticle Excitations 
around a Vortex Core in Quantum Limit}

\author{N. Hayashi,  T. Isoshima, M. Ichioka, and K. Machida}
\address{Department of Physics, Okayama University, Okayama 700, Japan}
\date{Submitted 26 September 1997}

\maketitle

\begin{abstract}
   Focusing on a quantum-limit behavior,
we study a single vortex in a clean $s$-wave type-II superconductor
by self-consistently solving
the Bogoliubov-de Gennes equation.
   The discrete energy levels of the vortex bound states
in the quantum limit is discussed.
   The vortex core radius shrinks monotonically up to an atomic-scale length
on lowering the temperature $T$, and the shrinkage stops to saturate
at a lower $T$.
   The pair potential, supercurrent, and
local density of states around the vortex exhibit Friedel-like
oscillations.
   The local density of states has particle-hole asymmetry
induced by the vortex.
   These are potentially observed directly by STM.
\end{abstract}


   Growing interest has been focused on vortices both 
in conventional and unconventional superconductors
from fundamental and applied physics points of view. 
   This is particularly true for high-$T_c$ cuprates,
since it is essential that one understands fundamental physical 
properties of the vortices in the compounds to better
control various superconducting characteristics
of some technological importance.
   Owing to the experimental developments, it is not
difficult to reach low temperatures of interest where
distinctive quantum effects associated with
the discretized energy levels of the vortex bound states
are expected to emerge.
   The quantum limit is realized at the temperature
where the thermal smearing is narrower
than the discrete bound state levels\cite{kp}:
$T/T_c \leq 1/(k_{\rm F}\xi_0)$  with 
$\xi_0$=$v_F / \Delta_0$ the coherence length 
($\Delta_0$ the gap at $T=0$)
and $k_{\rm F}$ ($v_F$) the Fermi
wave number (velocity).
   For example, in a typical layered type-II superconductor NbSe$_2$ with 
$T_c=7.2$~K and $k_{\rm F}\xi_0\sim 70$, the quantum limit is reached below
$T<50$~mK.
   As for the high-$T_c$ cuprates, the corresponding temperature is rather
high: $T<10$ K for YBa$_2$Cu$_3$O$_{7-\delta}$ (YBCO).

   Important microscopic works to theoretically investigate
the quasiparticle spectral structure around a vortex in a
clean limit are put forth by
Caroli {\it et al.}\cite{caroli}, 
Kramer and Pesch\cite{kp}, and
Gygi and Schl\"uter\cite{gygi}.
   The low-lying excitations are essential to correctly
describe low-$T$ thermodynamic and transport properties in
the vortex state (or the mixed state).
   These include anomalous electric or
thermal Hall conductivity\cite{ong} and 
mysterious observations of the quantum magnetic 
dHvA oscillations\cite{corcoran};
various topics are debated intensively\cite{otterlo}.
   Yet there has been no serious attempt or quantitative 
calculation to explore deep into the quantum 
regime.

   The purposes of the present paper
are to reveal the quantum-limit aspects of the single vortex
in $s$-wave superconductors
and to discuss a possibility for the observation of them.

   The present study is motivated by the following recent 
experimental and theoretical situations:
(1) The so-called Kramer and Pesch (KP) effect\cite{kp,gygi,volovik,ichioka};
a shrinkage of the core radius upon lowering $T$
(to be exact, an anomalous increase in the slope of the pair potential
at the vortex center at low $T$)
is now supported by some experiments\cite{doettinger}.
   The $T$ dependence of the core size is studied by $\mu$SR
on NbSe$_2$ and YBCO\cite{sonier},
which is discussed later.
   The KP effect, if confirmed, 
forces us radically alter the traditional 
picture\cite{fetter} for the vortex line 
such as a rigid normal cylindrical rod with the radius $\xi_0$.
(2) The scanning tunneling microscopy (STM) experiment on YBCO by
Maggio-Aprile {\it et al.}\cite{maggio}, which enables us to directly 
see the spatial structure of the low-lying quasiparticle excitations around
the vortex, arouses much interest.
   They claim that surprisingly enough, 
there exist only a few discretized bound-state levels in
the vortex core, i.e., the vortex is almost ``empty.''
   It resembles our naive image for conventional $s$-wave superconductors
where in the quantum limit a few quantized levels of the bound states
remain inside the bulk energy gap $\Delta_0$.
(3) The theoretical situation on this subject\cite{maggio}
is still very confusing;
   Some\cite{wang} claim that
the bound-state energy levels are
not discretized for $d$-wave pair,  
but discretized for $s$-wave pair.  
   Some\cite{morita} claim the discretized-like structure
even for the former.
   For $s$-wave case, where the formulation of the 
problem is well defined, we should establish our understanding
of the vortex structure in the quantum limit.
(4) Lastly, we are motivated by a curiosity;
   Previously we have calculated the local density of states (LDOS)
for $s$-wave pair on the basis of the quasiclassical (Eilenberger)
theory\cite{hayashi},
successfully applied to the STM observations on NbSe$_2$ done by Hess
{\it et al.}\cite{hess}.
   We are particularly interested in what happens in LDOS
at further lower $T$, say, below 50~mK deep into the
quantum limit, 
at which it may be now feasible to perform STM experiments.

   Prompted by these motivations,
we self-consistently solve the Bogoliubov-de Gennes (BdG)
equation, which is one of the most fundamental
microscopic equations of superconductivity and contains fully 
quantum effects.
   We start with the BdG equations
for the quasiparticle
wave functions $u_j({\bf r})$ and $v_j({\bf r})$
labeled by the quantum number $j$:
\begin{eqnarray}
\biggl[ {-1\over 2 k_{\rm F}\xi_0}\nabla^2-E_{\rm F} \biggr] u_j({\bf r})+
\Delta({\bf r})v_j({\bf r})
=E_j u_j({\bf r}),
\nonumber
\end{eqnarray}
\begin{eqnarray}
-\biggl[ {-1\over 2 k_{\rm F}\xi_0}\nabla^2-E_{\rm F} \biggr] v_j({\bf r})+
\Delta^{\ast}({\bf r})u_j({\bf r})
=E_j v_j({\bf r}),
\label{eq:1}
\end{eqnarray}
in a dimensionless form,
where $\Delta({\bf r})$ is the pair potential and
$E_{\rm F}$ (=$k_{\rm F}\xi_0 /2$) the Fermi energy.
   The length (energy) scale is measured by $\xi_0$ ($\Delta_0$).
   For an isolated single vortex
in an extreme type-II superconductor,
we may neglect the vector potential in Eq.\ (\ref{eq:1}).
   The pair potential is determined self-consistently
by
\begin{eqnarray}
\Delta({\bf r})=
g\sum_{|E_j|\leq \omega_{\rm D}}u_j({\bf r})v^{\ast}_j({\bf r})\{1-2f(E_j)\}
\label{eq:2}
\end{eqnarray}
with the Fermi function $f(E)$.
   Here, $g$ is the coupling constant and $\omega_{\rm D}$ the energy 
cutoff, which are related by a BCS relation
via the transition temperature $T_c$ and
the gap $\Delta_0$.
   We set $\omega_{\rm D}=10\Delta_0$.
   The current density is given by
${\bf j}({\bf r})$
$\propto {\rm Im} \sum_j\big[f(E_j)u^{\ast}_j({\bf r})\nabla u_j({\bf r})
+\{1-f(E_j)\}v_j({\bf r})\nabla v^{\ast}_j({\bf r}) \big]$.
   We consider an isolated vortex under the following conditions.
(a) The system is a cylinder with a radius $R$.
(b) The Fermi surface is cylindrical, appropriate for the materials such as
NbSe$_2$ and high-$T_c$ cuprates.
(c) The pairing has isotropic $s$-wave symmetry.
   Thus the system has a cylindrical symmetry.
   We write the eigenfunctions as
$u_j({\bf r})=u_{n\mu}(r) \exp\bigr[i(\mu-{1\over 2})\theta \bigl]$ and
$v_j({\bf r})=v_{n\mu}(r) \exp\bigr[i(\mu+{1\over 2})\theta \bigl]$ with
$\Delta({\bf r})=\Delta(r) \exp\bigr[-i\theta \bigl]$ 
in polar coordinates, where $n$ is a radial quantum number and 
the angular momentum $|\mu|={1\over 2},{3\over 2},{5\over 2},\cdots$.
   We expand the eigenfunctions in terms of the Bessel functions $J_m(r)$
as $u_{n\mu}(r)=\sum_ic_{ni}\phi_{i|\mu-{1\over2}|}(r)$ and $v_{n\mu}(r)=
\sum_id_{ni}\phi_{i|\mu+{1\over2}|}(r)$
with $\phi_{im}(r)={{\sqrt 2}\over RJ_{m+1}(\alpha_{im})}
J_m(\alpha_{im}r/ R)$,
$\bigl(i=1,2,\cdots, N$, and $\alpha_{im}$ is the $i$-th
zero of $J_m(r) \bigr)$.
   The BdG is reduced to a $2N\times 2N$
matrix eigenvalue problem\cite{gygi}.
   Our system is characterized by $k_{\rm F}\xi_0$,
which is a key parameter of the present problem.

   In Fig.\ \ref{fig:1}, the calculated spatial variation 
of $\Delta(r)$ is displayed
for various $T$. It is seen that as $T$ decreases, 
the core size $\xi_1$ defined by $\xi_1^{-1}=\lim_{r\rightarrow 0}
\Delta(r)/r$ shrinks and the oscillatory
spatial variation with a wave length $\sim 1/k_{\rm F}$
becomes evident in $\Delta(r)$\cite{kp,gygi}.
   The physical reason for this Friedel-like oscillation
lies in the following facts.
   All eigenfunctions $u_{n\mu}(r)$
and $v_{n\mu}(r)$ contain a rapid oscillation component
with $1/k_{\rm F}$.
   At lower $T$ the lowest 
bound states, whose oscillation amplitude is large
near the core, dominate physical quantities.
   We note that the oscillatory behavior can always appears at 
sufficiently low $T$ irrespective of values of $k_{\rm F}\xi_0$.
   We also mention that a similar oscillatory spatial
variation around a vortex core in  the Bose condensate
of $^4$He is found theoretically,
due to the roton excitations\cite{giorgini}.

   The associated supercurrent $j_{\theta}(r)$ and the field $H(r)$
are shown in Fig.\ \ref{fig:2}.
   Reflecting the above oscillation,
$j_{\theta}(r)$ also exhibits a weak oscillation around
$r=0.2$--0.5~$\xi_0$.
   It is difficult to see the oscillation in $H(r)$,
because it is obtained by
integrating $j_{\theta}(r)$ via the Maxwell equation
$\nabla\times{\bf H}={4\pi\over c}{\bf j}({\bf r})$,
resulting in a smeared profile.
   It is also seen that the position of the maximum of
$j_{\theta}(r)$ becomes shorter as $T$ decreases. 
   These features quite differ 
from those obtained within the Ginzburg-Landau (GL)
framework\cite{fetter,brandt}.

   The $T$ dependence of $\xi_1(T)$ for various $k_{\rm F}\xi_0$
values is shown in Fig.\ \ref{fig:3}.
   Coinciding with Kramer and Pesch\cite{kp} for  $s$-wave pair
and Ichioka {\it et al.}\cite{ichioka} for $d$-wave pair,
$\xi_1(T)$
decreases almost linearly with $T$, that is,
$\xi_1(T)/\xi_0 \sim T/ T_c$ except at extremely low $T$. 
   An important difference
from these quasiclassical theories\cite{kp,ichioka} 
appears at lower $T$.
   At a lower $T< T_0\simeq T_c/(k_{\rm F}\xi_0)$,
where the quantum limit is realized,
the shrinkage of the core size stops to saturate, and
   the saturated value is estimated as 
$\xi_1/ \xi_0 \sim (k_{\rm F}\xi_0)^{-1}$.

   According to the $\mu$SR experimental data\cite{sonier}, 
the core radius in NbSe$_2$ shows a strong $T$ dependence, 
while that in YBCO with $T_c$=60~K is almost $T$-independent
below $\sim$0.6$T_c$.
   This seemingly contradicting result can be understood
as follows.
   The strong $T$ dependence in NbSe$_2$ is the usual 
KP effect corresponding to the curves for larger $k_{\rm F}\xi_0$
in Fig.\ \ref{fig:3}.
   At lower $T$ than $T_0$ estimated as $\sim$100~mK
($k_{\rm F}\xi_0\sim$ 70),
the shrinkage must
saturate (the above experiment is done above $\sim$2~K). 
   As for the YBCO data, since the estimated $k_{\rm F}\xi_0$ is small
($\sim$4\cite{maggio} for YBCO with $T_c$=90~K),
the saturation is already attained at a relatively
high $T$ such as shown in Fig.\ \ref{fig:3}.
   Thus the absence or weakness of the KP effect in YBCO is simply
attributable to the fact
that the quantum-limit temperature $T_0$ is quite high.

   Reflecting the shrinkage of the core radius,
the bound-state energies $E_{\mu}$ increases as $T$ decreases.
   This $T$-dependent $E_{\mu}$ shift, due to the KP effect,
and its saturation at lower $T$
may lead to a nontrivial $T$ dependence in thermodynamic 
and transport properties.

   In Fig.\ \ref{fig:4}, we plot the energy levels $E_{\mu}$
of the low-lying bound states
$\bigl(\mu={1\over2},{3\over2},\cdots,{13 \over 2} \bigr)$
as a function of $k_{\rm F}\xi_0$,
at sufficiently low $T$ ($T/T_c=0.01$) where increasing of the energy levels
saturates.
   It is seen that in large-$k_{\rm F}\xi_0$
region, the bound states densely pack inside the gap $\Delta_0$, 
allowing us to regard them as continuous ones.
   This is the case
where the quasiclassical approximation\cite{kp,ichioka} validates.
   In small-$k_{\rm F}\xi_0$ region,
where the quantum effect is important even at high $T$,
only a few bound states remain within the low-energy region.
   We find that even in small-$|\mu|$ region,
the spacing between the energy levels $E_{\mu}$ is not constant,
but rather becomes narrower as $|\mu|$ increases.
   The often adopted formula
$E_{\mu}/ \Delta_0=2\mu/ (k_{\rm F}\xi_0)$ or
$2\mu/ (k_{\rm F}\xi_1)$
due to Caroli {\it et al.}\cite{caroli}, or
$E_{\mu}/ \Delta_0=(2\mu/k_{\rm F}\xi_0)\ln[\xi_0 / 2 \xi_1]$
by Kramer and Pesch in the limit $\xi_1 \ll \xi_0$\cite{kp}
do not satisfactorily explain our self-consistent results.
   Instead, our result is empirically fitted to a formula
$E_{1/2}/\Delta_0=
(0.5/ k_{\rm F}\xi_0) \ln[k_{\rm F}\xi_0 /0.3]$
for large $k_{\rm F}\xi_0$ as shown in the dotted curve in Fig.\ \ref{fig:4}.

   In Fig.\ \ref{fig:5}, the spectral evolution, i.e.,
the spatial variation of LDOS, which is calculated by 
$N({\bf r},E) \propto \sum_j \bigl[|u_j({\bf r})|^2 f'(E-E_j)
+|v_j({\bf r})|^2 f'(E+E_j) \bigr]$,
is shown for $k_{\rm F}\xi_0$=8 at low temperature $T$=0.05$T_c$.
   It is well contrasted with that of the higher $T$ case
by Gygi and Schl\"uter\cite{gygi} (see, for comparison,
 Fig. 15 in Ref.\ \cite{gygi} where $k_{\rm F}\xi_0 \sim 70$ and
$T \simeq 0.13 T_c$,
calculated under the two-dimensional Fermi surface).
   As lowering $T$, because of the quantum effects,
the thermally smeared spectral structure drastically changes and becomes
far finer one around the vortex.
   The spectra are discretized inside the gap and consist
of several isolated peaks, each of which precisely 
corresponds to the bound states
$E_{\mu}$ $\bigl( |\mu|={1\over2},{3\over2},\cdots \bigr)$.
   Reflecting the oscillatory nature of the eigenfunctions
$u_{\mu}(r)$ and $v_{\mu}(r)$
with the period $1/k_{\rm F}$, the spectral evolution also
exhibits the Friedel-like oscillation as seen from Fig.\ \ref{fig:5}.

   To show clearly the particle-hole asymmetry of
the LDOS of Fig.\ \ \ref{fig:5},
which is another salient feature,
we present in Fig.\ \ref{fig:6} 
the spectra at the vortex center $r=0$ and $0.2\xi_0$
$\bigl[$We can barely see the asymmetry in Wang and MacDonald\cite{wang}
$\bigl($see Fig. 3(a) in Ref.\ \cite{wang}$\bigr)\bigr]$.
   At the center $r=0$, the bound-state peak with $E_{1/2}$
appears only at $E>0$ side, because the eigenfunction 
$u_{1/2}(r=0)\neq 0$,
which consists of the Bessel function $J_0(r=0)$ $(\neq 0)$,
and all others for
$|E_{\mu}|<\Delta_0$ vanish at $r=0$.
   At $r=0.2\xi_0$, the other bound-state peaks are seen.
   The particle-hole asymmetry in the vortex bound states appears
even if the normal-state density of states is symmetric.
   These features are subtle\cite{gygi} or
absent\cite{hayashi} in the previous calculations.
   This asymmetry around the vortex is
quite distinctive, should be checked by STM experiments, and
may be crucial for the Hall conductivity in the mixed state.

   Let us argue some of the available experimental data 
in the light of the present study.
   The lowest bound state level
$E_{1/2}/ \Delta_0$ is estimated
by Maggio-Aprile {\it et al.}\cite{maggio}
for YBCO with $T_c$=90~K ($E_{1/2}$=5.5~meV and $\Delta_0$=20~meV),
yielding $k_{\rm F}\xi_0\sim 4$.
   Since it implies that $\xi_0$ is only of the order
of the crystal-lattice constant,
we should caution that Maggio-Aprile
{\it et al.}\cite{maggio} take their data for the spectral evolution every
10~\AA\ apart near the core, thus the important spatial information on LDOS 
might be lost.
   So far the existing STM data\cite{maggio,hess,dewilde} 
taken at the vortex center are almost symmetric about $E$=0,
e.g., on NbSe$_2$ at $T$=50~mK\cite{hess}.
   The reason why the so-called zero-bias peak is centered just
symmetrically at $E$=0
is that $k_{\rm F}\xi_0$ is large and $T$ is too
high to observe the quantum effects.

   We emphasize that in any clean $s$-wave type-II superconductors at 
appropriately low $T$
$\bigl( < T_0\simeq T_c/(k_{\rm F}\xi_0) \bigr)$,
one can observe these eminent characteristics associated with
the quantum effects.
   For example, a typical A-15 compound V$_3$Si\cite{corcoran}
with $T_c$=17~K, $\xi_0\simeq60$~\AA\ ($k_{\rm F}\xi_0 \simeq 12$),
and a borocarbide
LuNi$_2$B$_2$C\cite{dewilde} with $T_c$=16~K,
$\xi_0\simeq80$~\AA\ are the best candidates to check our results.

   In summary, we have analyzed the vortex core structure and the related 
quasiparticle energy spectrum by self-consistently solving
the BdG equation for an isolated vortex in a clean $s$-wave
type-II superconductor, focusing on the low-$T$ quantum effects.
   We have found the far richer
structure in the pair potential, supercurrent, and
LDOS than what
one naively imagines from the corresponding calculations
done at high $T$ or $k_{\rm F}\xi_0 \gg 1$\cite{gygi,hayashi},
and pointed out experimental feasibility to observe it.

   The widely used working hypothesis for
the vortex core of a rigid normal rod
with the radius $\xi_0$\cite{fetter}
must be cautiously used for the clean
superconductors of interest:
the magnetic field distribution probed by
neutron diffraction\cite{yethiraj} or
$\mu$SR\cite{sonier}
through the magnetic form factor analysis based on the GL theory
must be taken with caution.
   Detailed investigations of various mysteries
associated with the vortices,
e.g., the thermal Hall conductance\cite{ong} belong to future work.

We would like to thank J. E. Sonier and  A. Yaouanc for
useful discussions.
\begin{figure}
\caption{
     The spatial variation of the pair potential $\Delta(r)$
     normalized by $\Delta_0$ around the vortex for several temperatures
     and $k_{\rm F}\xi_0=16$.
     The length $r$ is measured by $\xi_0$.
}
\label{fig:1}
\end{figure}
\begin{figure}
\caption{
     The current distribution normalized by $c\phi_0/(8\pi^2\xi_0^3\kappa^2)$ 
     for several temperatures,
     where $\phi_0$ is the flux quantum and $\kappa$ ($\gg$1) is
     the GL parameter.
     The inset shows the field distribution normalized by
     $\phi_0/(2\pi\xi_0^2\kappa^2)$.
     The temperatures are the same as in Fig. 1, and $k_{\rm F}\xi_0=16$.
}
\label{fig:2}
\end{figure}
\begin{figure}
\caption{
     The $T$ dependence of the vortex radius $\xi_1$ normalized by $\xi_0$
     for several 
     $k_{\rm F}\xi_0$ ($=1.2$, 2, 4, and 16 from top to bottom).
}
\label{fig:3}
\end{figure}
\begin{figure}
\caption{
    The lowest seven bound-state energies $E_{\mu}$,
    normalized by $\Delta_0$,
    as a function of $k_{\rm F}\xi_0$,
    at enough low temperature $T/T_c=0.01$.
    The dotted line is a fitting curve (see the text).
}
\label{fig:4}
\end{figure}
\begin{figure}
\caption{
    The spectral evolution $N(E,r)$ at $T/T_c=0.05$ and $k_{\rm F}\xi_0=8$.
    It is normalized by 
    the normal-state density of states at the Fermi surface.
    $E$ and $r$ are measured by $\Delta_0$ and $\xi_0$, respectively.
}
\label{fig:5}
\end{figure}
\begin{figure}
\caption{
    The local density of states $N(E,r)$
    at $r=0$ (solid line) and $0.2\xi_0$ (dotted line).
    $T/T_c=0.05$ and $k_{\rm F}\xi_0=8$.
}
\label{fig:6}
\end{figure}

\end{document}